\documentclass[10pt,twocolumn]{article}
\usepackage{fix-cm} 
\usepackage[T1]{fontenc}
\usepackage[a4paper,margin=1in]{geometry}
\usepackage{booktabs}
\usepackage{amsmath,amssymb}
\usepackage{graphicx}
\usepackage{float}
\usepackage{xcolor}
\usepackage{draftwatermark}
\SetWatermarkText{PREPRINT}
\SetWatermarkScale{0.80}
\SetWatermarkColor[gray]{0.93}
\SetWatermarkAngle{45}
\usepackage{hyperref}
\usepackage{tabularx}
\usepackage{balance}
\usepackage{microtype}
\usepackage{listings}
\usepackage{fvextra}
\usepackage{placeins}
\newcommand{\safeincludegraphics}[2][]{\includegraphics[#1]{#2}}
\hypersetup{
  colorlinks=true,
  linkcolor=blue!55!black,
  urlcolor=blue!55!black,
  citecolor=blue!55!black,
  pdftitle={Portable Models as a Replacement for Industrial Heuristics in Compiler Optimizations},
  pdfauthor={Fot Nikolai and Vinarsky Alexander},
  pdfsubject={Portable function-inlining prediction from compiler diagnostics},
  pdfkeywords={compiler optimization, function inlining, machine learning, intermediate representation, portable compilers}
}

\title{Portable models as a replacement for industrial heuristics in compiler optimizations}
\author{%
\small
\begin{tabularx}{\dimexpr\textwidth-2\tabcolsep\relax}{@{}>{\centering\arraybackslash}X >{\centering\arraybackslash}X@{}}
\textbf{Fot Nikolai} & \textbf{Vinarsky Alexander} \\
Department of Compiler Technologies & Department of Compiler Technologies \\
Ivannikov Institute for System Programming (ISP) of the Russian Academy of Sciences &
Ivannikov Institute for System Programming (ISP) of the Russian Academy of Sciences \\
Moscow, Russia, 109125 & Dolgoprudny, Russia, 141701 \\
\texttt{ns.fot@ispras.ru} & \texttt{AlexanderVinarsky@yandex.ru}
\end{tabularx}%
}

\date{}
\begin{document}
\maketitle

\begin{abstract}
\par The paper investigates the possibility of predicting function-inlining decisions in compact compilers, source-to-source tools, and interpreters where the reuse of GCC or LLVM optimization infrastructure is impractical. The relevance of this work is determined by the need to transfer mature inlining heuristics to systems with limited compiler infrastructure, restricted runtime dependencies, and reduced access to target-specific analysis. Existing production compilers already contain strong inliners, but their decisions depend on internal intermediate representations (IRs), pass ordering, target models, and analysis stacks that are difficult to reproduce in lightweight systems.
\par To overcome these constraints, we propose a portable inlining-prediction framework. Production compiler diagnostics serve as supervision; a separate extractor reconstructs caller-callee callsites, prepares sterile source snippets, normalizes them into a universal AST, optionally lowers them to a lightweight structural IR, and exports scalar features for model training. Thus, a trained predictor can be emitted as ordinary C code without a compiler-runtime dependency.
\par To evaluate the proposed framework, we constructed a dataset comprising 336,938 callsites from fifteen open-source C projects, including 79,287 compiler-reported inline events. A comparison of several tabular models is performed using project-aware validation. Under leave-one-project-out validation, CatBoost reaches ROC-AUC 0.928 and PR-AUC 0.713; after threshold tuning, F1 improves from 0.670 to 0.729 and the false-positive rate drops from 0.192 to 0.084. Feature analysis shows that most signal is concentrated in source locality, explicit \texttt{inline} intent, callee size, side effects, branch and call structure, signature shape, and callsite argument shape.
\end{abstract}


\section{Introduction}

\par At present, function inlining remains one of the most important local optimizations in optimizing compilers. It replaces a function call with the body of the called function, removes direct call overhead, and may expose additional optimization opportunities to later passes, including constant propagation, dead-code elimination, scalar replacement, and loop optimizations. Its effect is both local and contextual: a profitable inline decision can improve the optimization context, whereas an unprofitable one can increase code size, register pressure, instruction-cache pressure, stack usage, and compilation time.
\par Production compilers manage this trade-off with mature cost models that depend on rich internal representations, target-specific information, and extensive engineering experience. However, these mechanisms are effective mainly inside their native compiler infrastructures. For small compilers, experimental frontends, educational compilers, interpreters, or language implementations without advanced intermediate representations, direct reuse of such mechanisms becomes difficult.
\par Therefore, this work studies a different deployment setting. The objective is to learn a compact inline-decision predictor from observed compiler behavior and to export it as a small, self-contained component. The model is not intended to improve GCC or LLVM directly. Instead, compiler diagnostics are used as supervision, and the learned predictor is intended for systems where only limited structural information about the caller, callee, and callsite is available.

\section{Problem Statement}

\par The target problem is binary callsite classification. Given the caller context, callee summary, source-local callsite information, and a small set of structural features, the model estimates whether a production-compiler-like inliner would inline the call. The output is an inline/no-inline hint, not a mandatory transformation. A downstream compiler may still apply conservative guards for unsupported constructs, recursion, large callee bodies, excessive stack growth, or code-size limits.
\par The main difficulty is caused by the portability requirement. Feature extraction must not require LLVM IR, GCC GIMPLE, tree-SSA, target-specific lowering, profile-guided metadata, alias analysis, or pass-manager state. Inference must be possible in a small runtime-free C component. The label source may be a production compiler, but the deployed predictor should be usable by systems closer to TinyCC/TCC, Smaller C, chibicc-like compilers, source-to-source tools, and interpreters with explicit function-call lowering hooks~\cite{tinyccProject,smallercProject,chibiccProject}.
\par Thus, the supervision used in this paper should be understood as compiler agreement. A positive label means that the reference compiler reported an inline event under the selected build configuration. A negative label means that the callsite was observed by the scanner but was not matched to such a report. These labels are useful for behavior distillation, but they should not be considered direct proof of runtime profitability.

\section{Existing Solutions}

\par Learning-based compiler optimization has several strong predecessors:

\begin{enumerate}
  \item \par \textit{MILEPOST GCC} demonstrates that static program features can support optimization choices within GCC~\cite{fursin2011milepost}. Its feature extractor uses GCC-internal representations such as tree-SSA, RTL, control-flow information, def-use chains, and loop information. Thus, it is an important reference point. Nonetheless, it is not a portable feature contract for small external compilers.
  \item \par \textit{MLGO} studies machine-learning-guided optimization in LLVM and includes inlining-for-size decisions~\cite{trofin2021mlgo}. MLGO operates within the LLVM optimization pipeline, uses LLVM IR and inliner state, and evaluates the production-compiler optimization metrics. The present work instead targets a predictor whose features are intentionally close to AST-level, file-level, and callsite-level structure, so that the inference path can later be reproduced outside LLVM. 
  \item \par \textit{CompilerGym} provides benchmark environments, action spaces, observation spaces, and reward interfaces for compiler-optimization research~\cite{cummins2022compilergym}. It is useful for experimentation, but it is not itself a compact inline heuristic that can be embedded in a small compiler. \textit{ProGraML} represents programs as attributed multigraphs with control-flow, data-flow, and call-flow edges~\cite{cummins2021programl}. Such representations are expressive, but graph construction and graph inference increase the deployment cost.
\end{enumerate}

\section{Architecture of the Proposed Solution}

\par
\begin{figure*}[t]
  \centering
  \safeincludegraphics[width=0.88\textwidth]{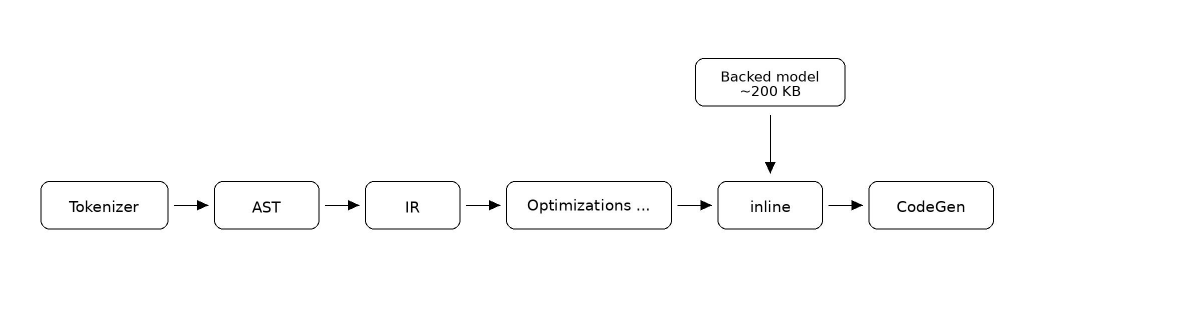}
  \caption{Integration strategy. The model consumes callsite and callee summaries and supplies an inline/no-inline hint before later compiler stages.}
  \label{img:pipeline-integration}
\end{figure*}

 Figure~\ref{img:pipeline-integration} summarizes the intended integration point. A frontend or small compiler supplies a callsite summary and a callee summary. The model returns an inline/no-inline hint before later lowering or backend stages. This placement is deliberately early: it does not require full SSA construction, register allocation, target-specific cost modeling, or profile-guided information.

\par This integration is supported by three main architectural layers. The first layer is scraping: rebuilding real projects with inline diagnostics, parsing positive events, enumerating candidate callsites, and reconstructing caller-callee pairs. The second layer is representation: preparing sterile snippets, normalizing them into a universal AST, lowering them into a lightweight structural IR when possible, and emitting scalar features. The third layer is modeling: training a baseline model, comparing model families, tuning thresholds and feature sets, and exporting the selected predictor as a plain C evaluator.

\subsection{Scraping Layer}

\par The scraping layer obtains labels from compiler feedback. It is intentionally independent of the model-training code: the same collected table can be used for CatBoost, LightGBM, random forests, logistic baselines, ablations, or a later handcrafted distilled tree. The diagnostic log is not sufficient by itself, because it gives positive events but not negative candidates, function bodies, signature information, or local callsite structure.

\subsection{Baseline Model Layer}

\par The baseline layer converts each labeled callsite into a feature vector and trains a deployable tabular model. Early proof-of-concept export used a shallow in-repository tree ensemble. The current reference model is CatBoost because it provides strong project-wise ranking quality and consists of oblivious trees that can be represented as static arrays. Random forests and Extra Trees remain important baselines because they are simple tree ensembles and can also be exported as standalone code.

\subsection{Tuning Layer}

\par Tuning is performed at two levels. Model-family tuning selects between linear, forest, and boosted-tree models. Feature-contract tuning compares compact subsets such as \texttt{core4}, \texttt{core8}, \texttt{core12}, \texttt{core20}, \texttt{ast\_site\_no\_ir}, and the full feature set. Decision-threshold tuning is separate from model training: a threshold can trade recall for a lower false-positive rate without changing the learned ranking function.

\section{Scraping}

\par The dataset is built by scraping compiler feedback. Each target project is rebuilt with inline optimization diagnostics enabled. A diagnostic line gives a positive label when the compiler reports that a particular callee was inlined into a caller. The simplified form of the two supported diagnostic styles is shown in Figure~\ref{img:gcc-inline-diagnostic}.

\begin{figure*}[t]
\centering
  \begin{Verbatim}[breaklines=true, breakanywhere=true, fontsize=\footnotesize]
  (GCC) source.c:123:123: optimized: Inlined callee.callee.123/123 into caller.caller.123/123 which now has time X and size Y, net change of -Z.

  (Clang) source.c:123:123: remark: 'callee' inlined into 'caller' with (cost=X, threshold=Y) at callsite caller:123:123; [-Rpass=inline]
  \end{Verbatim}
  \caption{Example GCC and Clang inline optimization diagnostics used as positive label sources.}
  \label{img:gcc-inline-diagnostic}
\end{figure*}

\par
\begin{figure*}[t]
  \centering
  \safeincludegraphics[width=0.95\textwidth]{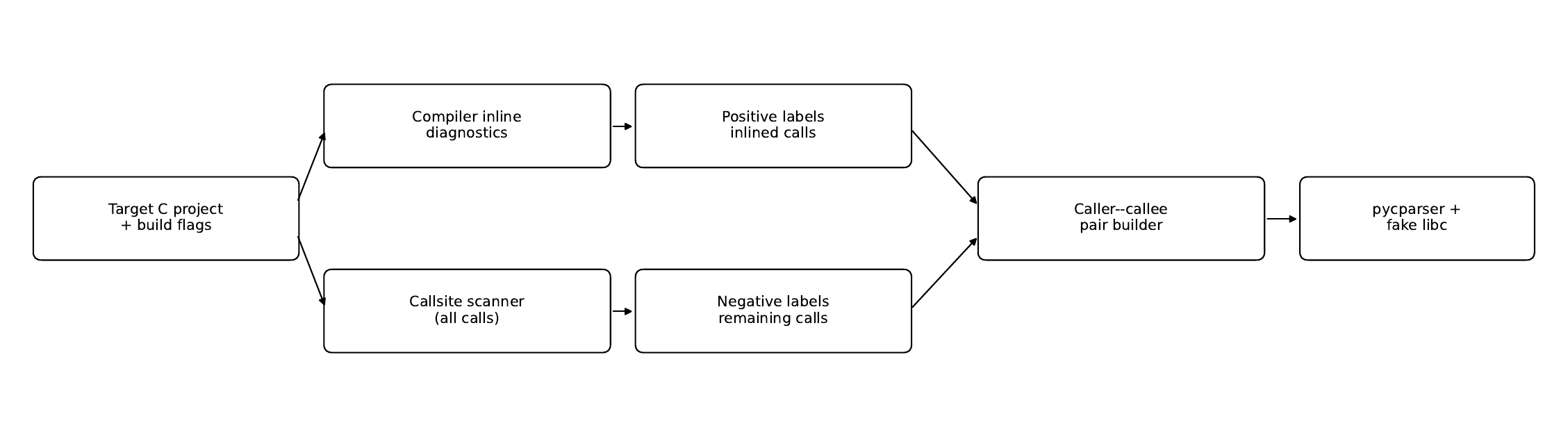}
  \caption{Scraping framework. Positive labels are extracted from compiler inline diagnostics. Negative labels are obtained from remaining callsites after removing known inline events. Both labeled paths are parsed and lowered into a normalized AST and a structural IR before feature extraction.}
  \label{img:scraping-framework}
\end{figure*}

 Figure~\ref{img:scraping-framework} shows the current scraping framework. The positive-label parser records the source location, caller, callee, and diagnostic text when these fields are available. A separate source-level scanner then collects callsites from the same project. A callsite is removed from the negative pool only when the complete key is matched: the normalized file, line, column, and the caller and callee names. This conservative policy avoids name-only matching across unrelated callsites.

\par After labeling, the extractor reconstructs a local caller-callee pair. It first looks in the event file and function index, then uses bounded fallback source search when necessary. Repeated diagnostics with the same file, caller, and callee are disambiguated by occurrence offset. The corresponding call occurrence inside the caller is then selected for feature extraction.
\par The output of this stage is a model-ready table. Each row represents one callsite. A positive row means that the reference compiler reported an inline event under the selected build configuration. A negative row means that the scanner found the callsite, but it was not matched to such a report. The negative pool can therefore contain unsupported, optimized-away, indirectly called, missed, or merely unselected calls.

\section{NeuralIR Pipeline}

\par The representation has two layers. The first is a normalized abstract syntax tree. The second is a lightweight structural intermediate representation. This split is the main reason why the framework is not tied to GCC or LLVM.
\par The normalized AST is meant to be a common source-level vocabulary. It stores only constructs that are useful for inline prediction: functions, calls, returns, loops, switches, branches, declarations, assignments, normalized unary and binary operations, local objects, and visible side-effect patterns. It discards parser-specific details and many project-specific constructs.
\par
\begin{figure}[H]
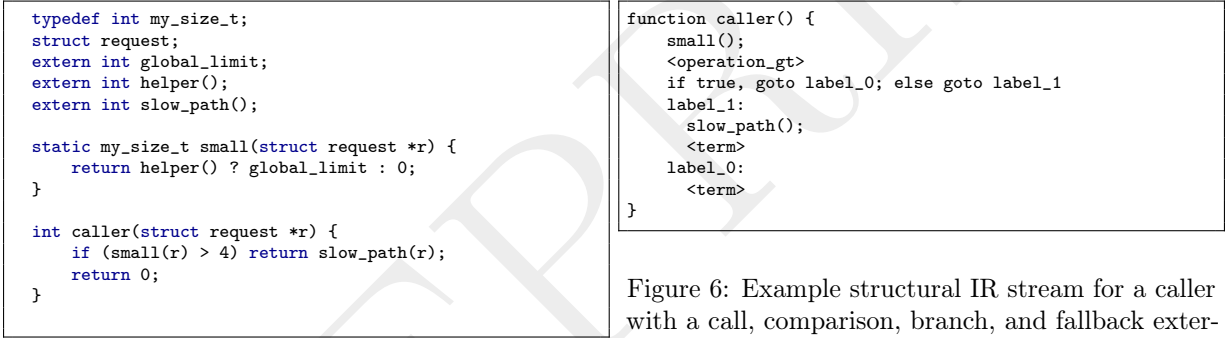

\centering
  \begin{lstlisting}[language=C]
  typedef int my_size_t;
  struct request;
  extern int global_limit;
  extern int helper();
  extern int slow_path();

  static my_size_t small(struct request *r) {
      return helper() ? global_limit : 0;
  }

  int caller(struct request *r) {
      if (small(r) > 4) return slow_path(r);
      return 0;
  }
  \end{lstlisting}
  \caption{Example C pair after preparation. The snippet keeps caller-callee structure but removes constructs that are not needed by the local predictor.}
  \label{img:c-after-cleaning}
\end{figure}

 Before a snippet is converted to the normalized AST, the extractor builds a minimal caller-callee translation unit. The goal is not to preserve the whole project, but rather the local inline decision context. Figures~\ref{img:c-before-cleaning} and~\ref{img:c-after-cleaning} show this preparation step.

\begin{figure}[H]
\centering
  \begin{lstlisting}[language=C]
  /* Comment. Comment. Comment. */
  #define likely(x) __builtin_expect(!!(x), 1)
  typedef unsigned long my_size_t;
  struct request { char *buf; my_size_t len; };
  static __attribute__((always_inline)) my_size_t small(struct request *r) {
      asm volatile("" ::: "memory");
      return likely(r->len) ? helper(r->buf, ({ my_size_t n = r->len; n; })) : global_limit;
  }
  int caller(struct request *r) {
      if (small(r) > 4) return slow_path(r);
      return 0;
  }
  \end{lstlisting}
  \caption{Example C pair before sterile snippet preparation.}
  \label{img:c-before-cleaning}
\end{figure}

\par The structural IR is not a production compiler IR and is not designed for code generation. It intentionally avoids SSA values, full type propagation, alias analysis, register allocation, target lowering, and rich data-flow edges. Instead, it uses a short action vocabulary over at most three operands per instruction. Function definitions become \texttt{fdecl}/\texttt{fend}; calls become \texttt{fcall} or \texttt{scall}; returns become \texttt{term}; branches and loops introduce labels and jumps; binary and unary expressions become operation counters.

\begin{figure}[H]
\centering
\begin{lstlisting}
function caller() {
    small();
    <operation_gt>
    if true, goto label_0; else goto label_1
    label_1:
      slow_path();
      <term>
    label_0:
      <term>
}
\end{lstlisting}
\caption{Example structural IR stream for a caller with a call, comparison, branch, and fallback external call.}
\label{img:structural-ir-stream-example}
\end{figure}

\par This intermediate layer exists for feature stability. A pure AST can answer many questions, but some structural questions are easier after lowering: how many operation-like actions exist, whether the callsite has local lowered context, how large a simplified basic-block structure is, and whether loop-local context surrounds the call. At the same time, the IR remains much cheaper than LLVM IR or GCC GIMPLE/RTL.

\begin{figure*}[t]
  \centering
  \safeincludegraphics[width=0.92\textwidth]{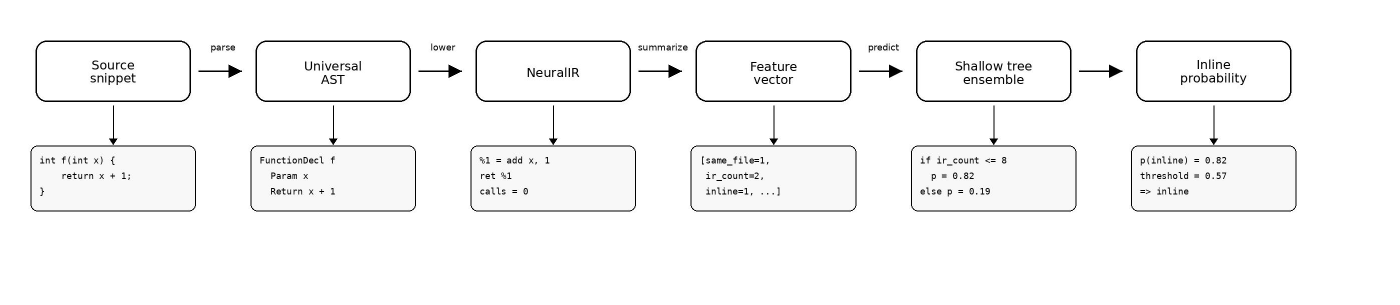}
  \caption{Portable inline-prediction architecture. Source snippets are parsed into a normalized AST, lowered to structural IR, summarized into a compact numeric feature vector, and evaluated by a tree ensemble.}
  \label{img:model-architecture}
\end{figure*}

\section{Basic Dataset Analysis}

\par The current corpus contains 336,938 callsites from fifteen C project scrapes: cJSON 1.7.19~\cite{cjsonProject}, FFmpeg 60.32.100~\cite{ffmpegProject}, Git 2.54.0~\cite{gitProject}, json-c 0.18.99~\cite{jsoncProject}, libevent 2.2.1-alpha-dev~\cite{libeventProject}, libuv 1.52.2-dev~\cite{libuvProject}, Linux~\cite{linuxProject}, Nginx 1.31.1~\cite{nginxProject}, PostgreSQL~\cite{postgresqlProject}, Redis 8.8.0~\cite{redisProject}, SQLite 3.54.0~\cite{sqliteProject}, tmux next-3.7~\cite{tmuxProject}, Valkey 9.1.0~\cite{valkeyProject}, Vim 9.2.286~\cite{vimProject}, and Zstd 1.6.0~\cite{zstdProject}. Overall, 79,287 callsites are labeled as compiler-inlined, giving a positive rate of 23.53\%.

\begin{table*}[t]
\centering
\caption{Multi-project callsite corpus after deduplication.}
\label{tab:dataset}
\scriptsize
\begin{tabular}{lrrrrr}
  \toprule
  Project & Callsites & Inlined & Inline, \% & Explicit inline & Explicit share, \% \\
  \midrule
  cJSON      & 367     & 0      & 0.00  & 0      & 0.00 \\
  FFmpeg     & 54,819  & 14,541 & 26.53 & 19,977 & 36.44 \\
  Git        & 58,221  & 6,946  & 11.93 & 15,251 & 26.20 \\
  JSONC      & 91      & 0      & 0.00  & 7      & 7.69 \\
  Libevent   & 1,889   & 0      & 0.00  & 111    & 5.88 \\
  libuv      & 179     & 0      & 0.00  & 41     & 22.91 \\
  Linux      & 63,786  & 27,641 & 43.33 & 44,527 & 69.81 \\
  Nginx      & 3,280   & 225    & 6.86  & 208    & 6.34 \\
  PostgreSQL & 119,967 & 22,421 & 18.69 & 30,911 & 25.77 \\
  Redis      & 310     & 99     & 31.94 & 50     & 16.13 \\
  SQLite     & 12,571  & 3,628  & 28.86 & 2      & 0.02 \\
  tmux       & 4,206   & 431    & 10.25 & 0      & 0.00 \\
  Valkey     & 11,976  & 1,299  & 10.85 & 1,321  & 11.03 \\
  Vim        & 311     & 72     & 23.15 & 0      & 0.00 \\
  Zstd       & 4,965   & 1,984  & 39.96 & 2,184  & 43.99 \\
  \midrule
  Total      & 336,938 & 79,287 & 23.53 & 114,590 & 34.01 \\
  \bottomrule
\end{tabular}
\end{table*}

\par
\begin{figure}[H]
  \centering
  \safeincludegraphics[width=\columnwidth]{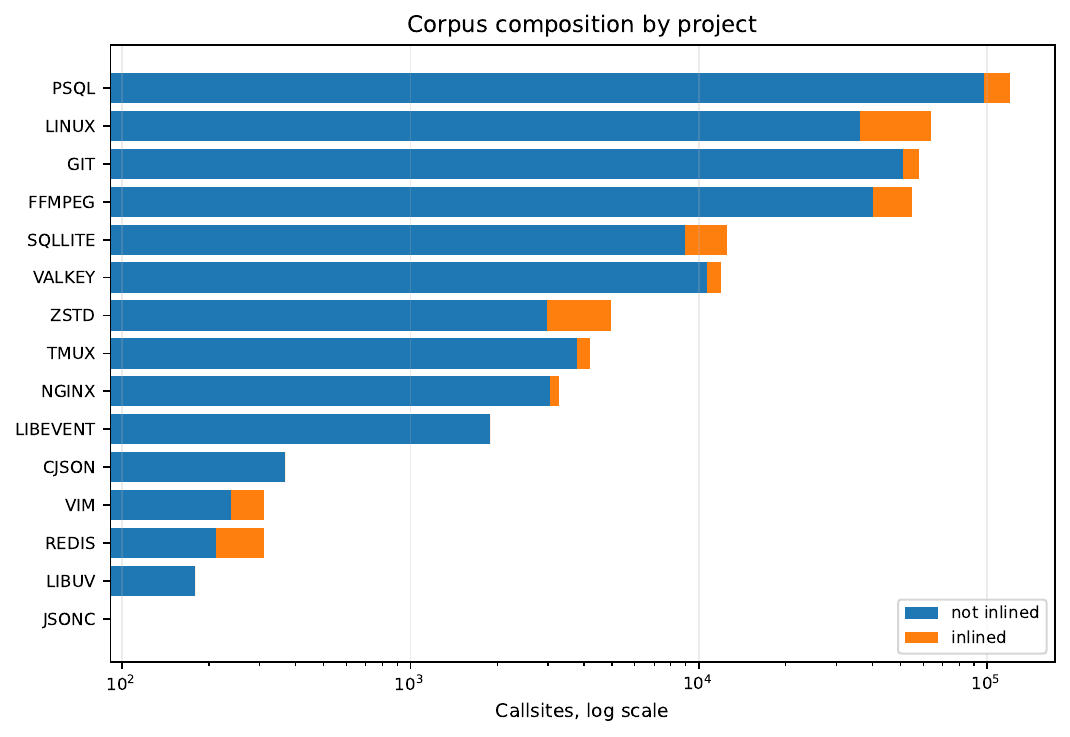}
  \caption{Corpus composition by project. The largest projects dominate row count, so random splitting is not enough to evaluate portability.}
  \label{img:corpus-by-project}
\end{figure}

\begin{figure}[H]
  \centering
  \safeincludegraphics[width=\columnwidth]{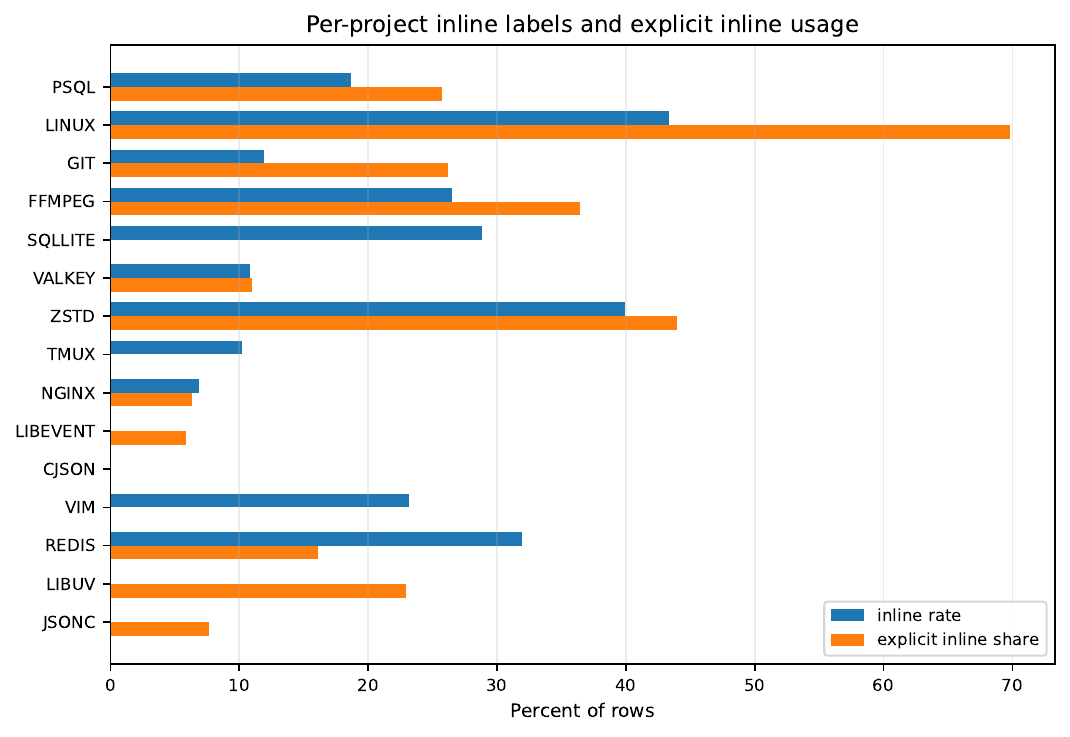}
  \caption{Per-project inline label rate and explicit-\texttt{inline} usage. Inline-label density varies substantially across projects.}
  \label{img:project-rates}
\end{figure}

 Figures~\ref{img:corpus-by-project} and~\ref{img:project-rates} show why project-aware validation is required. The corpus is large, but it is not balanced. PostgreSQL, Linux, Git, and FFmpeg dominate the row count; several smaller projects currently have no positive inline labels.

\par The explicit-inline feature captures programmer intent. Among 114,590 explicitly marked callsites, 64,322 are actually inlined by the compiler, giving an inline rate of 56.1\%. Among 222,348 callsites without the modifier, 14,965 are inlined, giving an inline rate of 6.7\%. Thus, explicit \texttt{inline} corresponds to a 49.4 percentage-point increase and an 8.34$\times$ relative lift.

\begin{figure}[H]
  \centering
  \safeincludegraphics[width=\columnwidth]{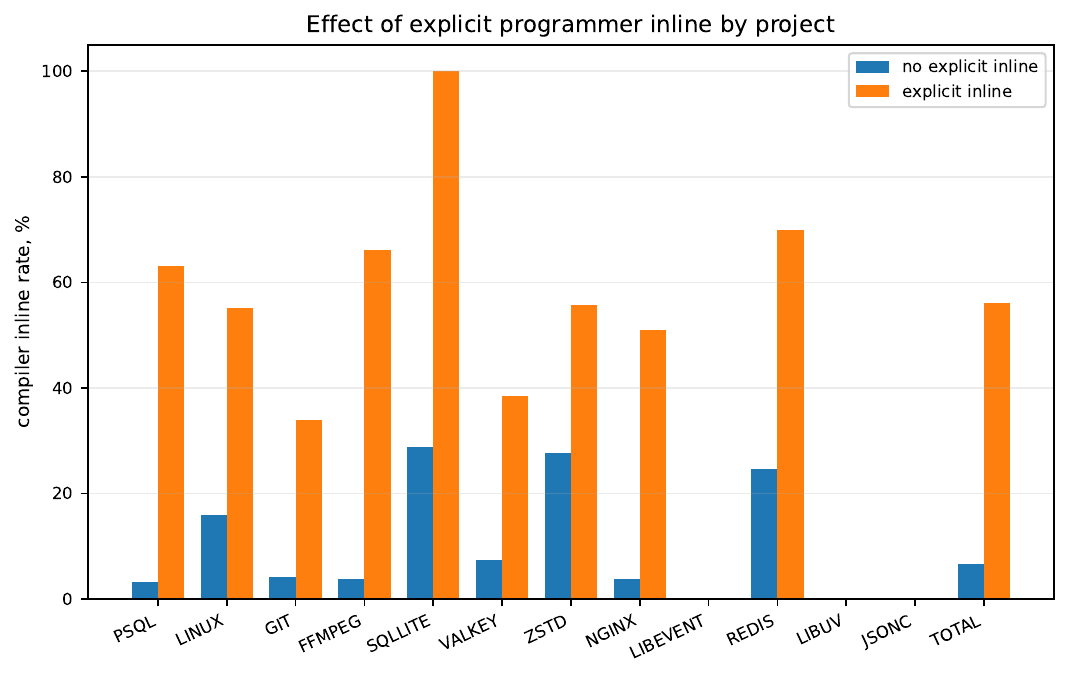}
  \caption{Compiler inline rate for explicitly marked and non-explicit callsites. Only projects with explicit-inline rows are shown, plus the total.}
  \label{img:inline-modifier}
\end{figure}

\section{Choosing the Optimal Model}

\par The full feature space contains 64 numeric and Boolean features. They describe callsite argument structure, callee signature, explicit \texttt{inline} modifiers, AST and IR complexity, side effects, local variables and arrays, operation counts, file locality, and loop context. The full representation is useful for analysis, but not every column should become part of a portable contract. Some features are unstable across projects, such as names and paths. Others duplicate the same compiler-cost signal.
\par The current minimal useful core is an 8-feature contract: same source file, explicit inline intent, side effects, function size, callee call count, branch count, parameter count, and call argument count. This set intentionally removes function names, file names, project names, and other memorization-prone identifiers.

\begin{table*}[t]
\centering
\caption{Feature-set ablation on the current multi-project corpus. All rows use the same source-aware split, source-class weighting, seed, and compact tree configuration: 24 trees, depth 5, and minimum leaf size 6.}
\label{tab:feature-sets}
\begin{tabular}{lrrrrrrr}
\toprule
Set & Feat. & Precision & Recall & F1 & FP & FN & Accuracy \\
\midrule
\texttt{core4} & 4 & 0.641 & 0.739 & 0.686 & 3,603 & 2,271 & 0.866 \\
\texttt{core8} & 8 & \textbf{0.676} & 0.874 & \textbf{0.762} & 3,647 & 1,093 & \textbf{0.891} \\
\texttt{core12} & 12 & 0.645 & \textbf{0.918} & 0.758 & 4,384 & \textbf{717} & 0.883 \\
\texttt{core20} & 20 & 0.672 & 0.849 & 0.750 & \textbf{3,602} & 1,314 & 0.887 \\
\texttt{all} & 64 & 0.659 & 0.881 & 0.754 & 3,963 & 1,036 & 0.886 \\
\bottomrule
\end{tabular}
\end{table*}

\begin{figure}[H]
  \centering
  \safeincludegraphics[width=\columnwidth]{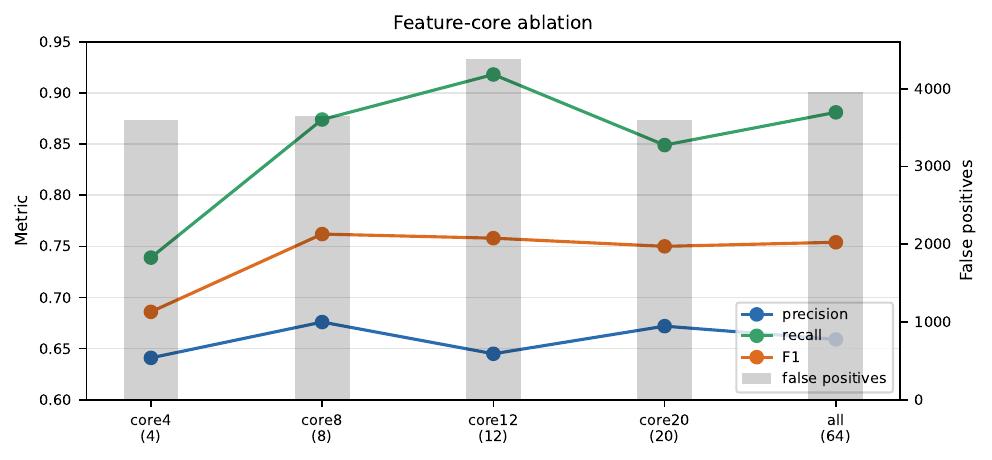}
  \caption{Feature-core ablation. The 8-feature core gives the best compact F1 and accuracy in the fixed-configuration comparison, while the 12-feature extension shifts the trade-off toward recall and more false positives.}
  \label{img:feature-set-comparison}
\end{figure}

\par Several tabular classifiers were considered: linear models, decision trees, random forests, Extra Trees, gradient boosting, XGBoost, LightGBM, and CatBoost. CatBoost is used as the current generated-C reference model because it gives strong project-wise ranking quality and can still be represented as static decision-tree tables.

\begin{table*}[t]
\centering
\caption{Comparison of evaluated model families under leave-one-project-out validation. \textit{F1*}, \textit{F0.5*}, and \textit{Acc.*} use the best threshold selected on the validation predictions.}
\label{tab:model-family-comparison}
\scriptsize
  \setlength{\tabcolsep}{3pt}%
  \begin{tabular}{lrrrrrrl}
  \toprule
  Model & Precision* & Recall* & F1* & F0.5* & FPR* & Acc.* & Deployment note \\
  \midrule
  LightGBM + negative weight~\cite{ke2017lightgbm} & 0.6423 & \textbf{0.773} & \textbf{0.702} & \textbf{0.665} & 0.133 & \textbf{0.845} & best F1, more false positives \\
  CatBoost~\cite{prokhorenkova2018catboost} & 0.661 & 0.667 & 0.664 & 0.663 & 0.105 & 0.841 & generated-C reference \\
  XGBoost + negative weight~\cite{chen2016xgboost} & 0.665 & 0.629 & 0.646 & 0.657 & 0.098 & 0.838 & low-FPR compromise \\
  LightGBM~\cite{ke2017lightgbm} & \textbf{0.679} & 0.576 & 0.623 & 0.656 & \textbf{0.084} & 0.836 & conservative tuned variant \\
  XGBoost~\cite{chen2016xgboost} & 0.661 & 0.634 & 0.647 & 0.656 & 0.100 & 0.837 & strong boosting baseline \\
  CatBoost + negative weight~\cite{prokhorenkova2018catboost} & 0.670 & 0.597 & 0.632 & 0.654 & 0.090 & 0.836 & conservative CatBoost variant \\
  Extra Trees~\cite{geurts2006extratrees} & 0.622 & 0.732 & 0.673 & 0.641 & 0.137 & 0.832 & randomized-forest baseline \\
  Random Forest~\cite{breiman2001randomforests} & 0.619 & 0.723 & 0.667 & 0.637 & 0.137 & 0.830 & portable ensemble baseline \\
  SGD Logistic~\cite{pedregosa2011scikit} & 0.607 & 0.712 & 0.655 & 0.625 & 0.142 & 0.824 & weak linear baseline \\
  \bottomrule
  \end{tabular}
\end{table*}

\begin{figure}[H]
  \centering
  \safeincludegraphics[width=\columnwidth]{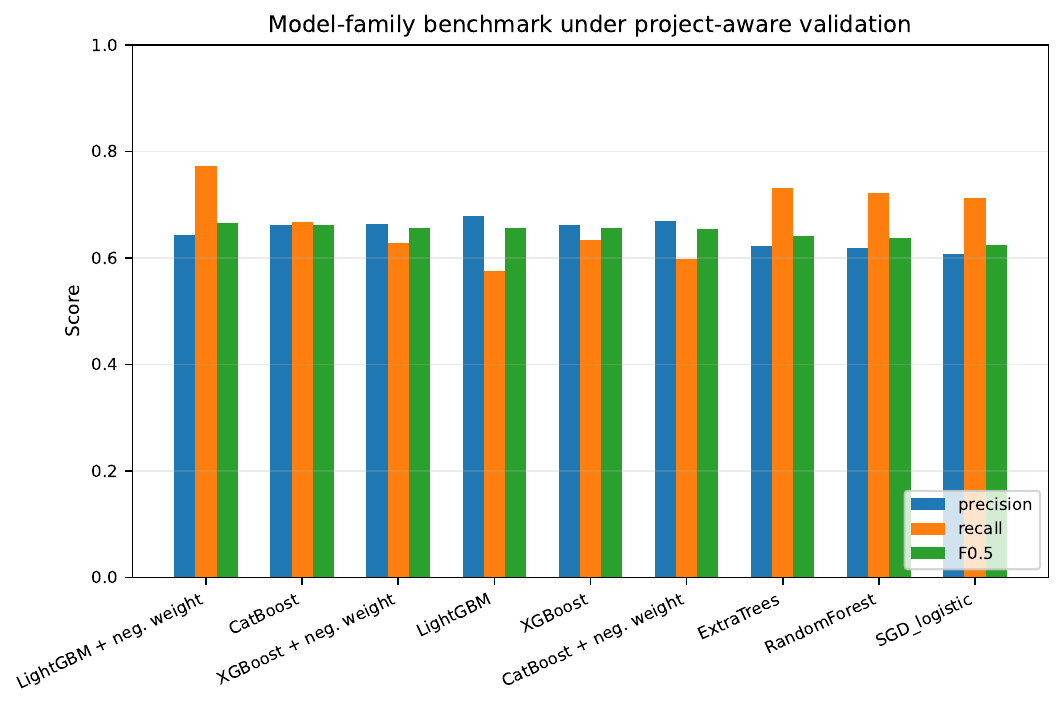}
  \caption{Offline model-family benchmark. Boosted-tree methods provide the strongest ranking metrics; CatBoost is used as the generated-C deployment candidate.}
  \label{img:model-family-benchmark}
\end{figure}

\section{Training}

\par The training procedure reads scraper CSV rows, normalizes the selected feature contract, and evaluates project-aware splits. Because projects differ strongly in size and positive rate, source-aware splitting and source-class weighting are preferred over random splitting. The main validation protocol is leave-one-project-out: each project is held out as the test set once, while the model is trained on the remaining projects.
\par The CatBoost deployment candidate uses 100 oblivious trees of depth 6 with learning rate 0.06 and L2 regularization 3.0. The tree structure is exported to JSON and then converted into a plain C evaluator. This keeps the same integration idea---a small compiler pass supplies scalar features and receives an inline/no-inline hint-but replaces the previous shallow export with the best verified model from the model-family benchmark.
\par CatBoost is suitable for this export because all nodes at the same depth in an oblivious tree use the corresponding split~\cite{prokhorenkova2018catboost}. Inference therefore reduces to computing a leaf index from a fixed sequence of feature-threshold comparisons and adding the selected leaf value. The generated C file contains no classes, no C++ standard library containers, no dynamic allocation, and no CatBoost runtime dependency.

\section{Evaluation of the Model}

\par The current experiments evaluate agreement with compiler inline diagnostics rather than downstream runtime or code-size improvement. At the default probability threshold of 0.5, CatBoost obtains ROC-AUC 0.9283, PR-AUC 0.713, and F1 0.670. The default threshold favors recall: it captures most compiler-inlined cases but produces many false positives. After tuning the decision threshold to 0.764, F1 improves to 0.729 and the false-positive rate drops from 0.192 to 0.084. This tuned threshold is the one embedded in the generated C predictor.

\begin{table}[H]
\centering
\caption{CatBoost confusion matrix under leave-one-project-out validation.}
\label{tab:catboost-confusion}
\tiny
  \setlength{\tabcolsep}{2pt}%
  \begin{tabular}{lrrrrrr}
  \toprule
  Threshold & TN & FP & FN & TP & F1 & FPR \\
  \midrule
  0.500 & 113,125 & 26,813 & 3,744 & 31,055 & 0.6702 & 0.1916 \\
  0.764 & 128,174 & 11,764 & 8,117 & 26,682 & 0.7286 & 0.0841 \\
  \bottomrule
  \end{tabular}
\end{table}

\par A separate CatBoost experiment tests the AST/site-only feature contract. It uses a lightweight leave-one-project-out setup and removes all structural-IR-derived counters. Despite this restriction, \texttt{ast\_site\_no\_ir} remains close to the full 64-feature CatBoost benchmark: ROC-AUC drops from 0.928 to 0.924, PR-AUC from 0.713 to 0.694, and F1 from 0.670 to 0.646. This is important because the AST/site contract is easier to reproduce outside GCC and LLVM.

\begin{table}[H]
\centering
\caption{AST/site-only CatBoost experiment without structural-IR-derived features. The negative-only FP rate is measured on projects/folds that contain no positive inline labels.}
\label{tab:ast-site-no-ir}
\tiny
  \begin{tabular}{lrrrrr}
  \toprule
  Set & Feat. & ROC & PR & F1 & Neg.-only FP \\
  \midrule
  \texttt{ast\_site\_no\_ir} & 42 & 0.924 & 0.694 & 0.646 & 0.146 \\
  Full CatBoost & 64 & 0.928 & 0.713 & 0.670 & -- \\
  \bottomrule
  \end{tabular}
\end{table}

\par Single-feature correlations are not a complete explanation of a tree ensemble, but they are useful for sanity checking. Larger AST depth, more branches per basic block, more parameters, more call arguments, more local variables, loops, assignments, comparisons, logic, and arrays are negatively associated with inlining. Leaf callees and compact return-expression structure are positively associated.

\begin{figure}[H]
  \centering
  \safeincludegraphics[width=\columnwidth]{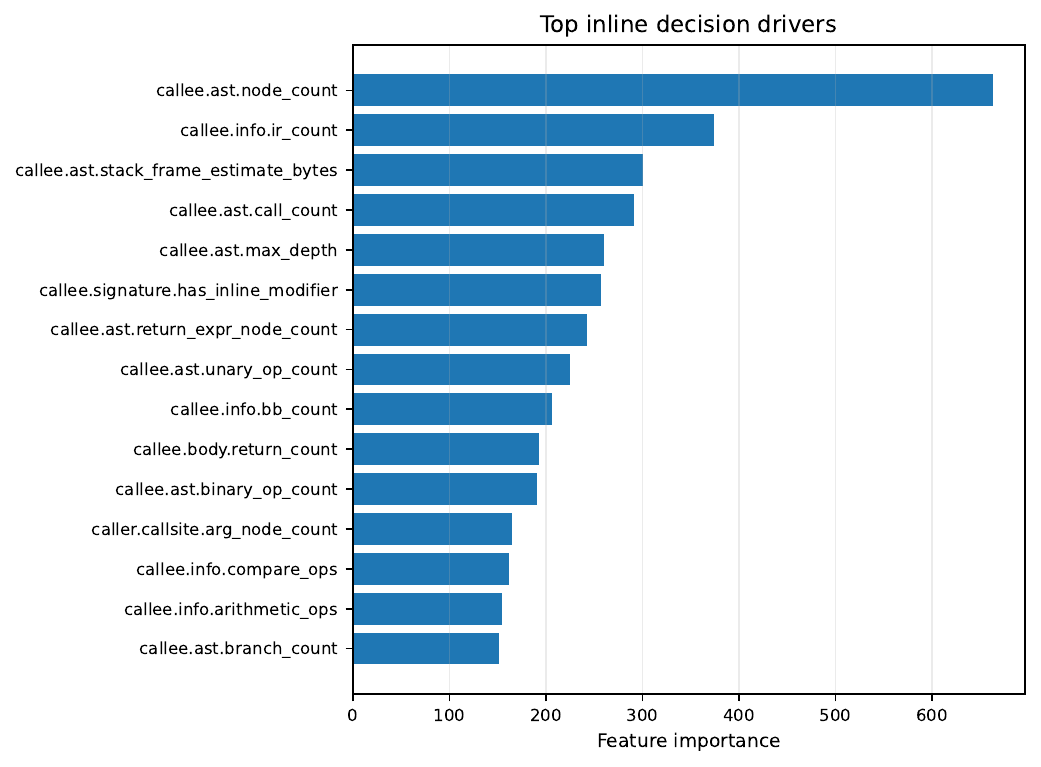}
  \caption{Top single-feature correlations with the inline label. Structural simplicity dominates the observed decision surface.}
  \label{img:inline-drivers}
\end{figure}



\FloatBarrier

\section{Tuning}

\par Tuning has two practical goals. The first is to choose a model that ranks positive callsites well under project-aware validation. The second is to select a threshold that makes the exported predictor usable in a compiler-like setting. A low threshold may maximize recall but can produce too many harmful false positives. A high threshold misses more potential inlines but is safer for downstream integration.
\par The current tuned threshold, 0.764, is selected from validation predictions and embedded in the generated C predictor. The threshold is intentionally treated as a deployment parameter rather than as a property of the feature extractor. Future tuning should be repeated after adding Clang labels, changing the project mix, or moving from compiler-agreement metrics to downstream runtime and code-size metrics.
\par The roadmap for tuning is to compare \texttt{core8}, \texttt{core12}, \texttt{ast\_site\_no\_ir}, and \texttt{all} under CatBoost and under smaller distilled-tree variants with stricter threshold policies that reduce false positives. Thematic ablations should isolate locality, explicit programmer intent, callee size, side effects, and callsite argument shape. 




\section{Conclusion}

\par This report presents a framework-centered baseline for portable function-inlining prediction. The pipeline scrapes compiler inline diagnostics, reconstructs labeled caller--callee callsites, prepares sterile source snippets, normalizes them into a universal AST, optionally lowers them to a lightweight structural IR, extracts scalar features, compares deployable tabular models, and exports the selected CatBoost model as plain C.
\par The current multi-project corpus contains 336,938 callsites, of which 23.53\% are labeled as compiler-inlined. The main empirical result is that nonlinear tree-based models are well suited to the representation. CatBoost gives ROC-AUC 0.928, PR-AUC 0.713, and tuned F1 0.7286, while LightGBM with negative-class weighting gives the highest tuned F1 in the model-family sweep. CatBoost is kept as the generated-C reference because its oblivious trees compile to a compact runtime-free evaluator.
\par The main representation result is that most signal is concentrated in locality, explicit inline annotations, callee size, side effects, branch and call structure, signature shape, and callsite argument shape. The AST/site-only experiment suggests that a useful lower-dependency model is possible even without structural-IR-derived counters.
\par The current evidence supports cross-project agreement with compiler diagnostics, not end-to-end improvement in runtime or binary size. The corpus is imbalanced, negative labels are noisy, and the representation intentionally avoids non-portable names, paths, and compiler-internal state. Therefore, the next version should prioritize compiler integration, benchmark measurements, and validation under additional label sources such as Clang-style optimization remarks. If the AST/site feature contract remains stable under these conditions, it can become the primary candidate for cross-language experiments on C/C++, Go, Lua, Python, and later additional compiled-language frontends.

\bibliographystyle{plain}
\bibliography{links}
\end{document}